# A CLOUD-NATIVE GLOBALLY DISTRIBUTED FINANCIAL EXCHANGE SIMULATOR FOR STUDYING REAL-WORLD TRADING-LATENCY ISSUES AT PLANETARY SCALE

**Bradley Miles[a] and Dave Cliff[b]**

Department of Computer Science
University of Bristol, Bristol, BS8 1UB, U.K.

[a]bm15731.2015@my.bristol.ac.uk, [b]csdtc@bristol.ac.uk,

**ABSTRACT**

We describe a new public-domain open-source simulator of an electronic financial exchange, and of the traders that interact with the exchange, which is a truly distributed and cloud-native system that been designed to run on widely available commercial cloud-computing services, and in which various components can be placed in specified geographic regions around the world, thereby enabling the study of planetary-scale latencies in contemporary automated trading systems. The speed at which a trader can react to changes in the market is a key concern in current financial markets but is difficult to study latency issues using conventional market simulators, and is extremely difficult to study "in the wild" because of the financial and regulatory barriers to entry in conducting experimental work on real financial exchanges. Our simulator allows an exchange server to be launched in the cloud, specifying a particular geographic zone for the cloud hosting service; automated-trading clients which attach to the exchange can then also be launched in the cloud, in the same geographic zone and/or in different zones anywhere else on the planet, and those clients are then subject to the real-world latencies introduced by planetary-scale cloud communication interconnections. In this paper we describe the design and implementation of our simulator, called DBSE, which is based on a previous public-domain simulator, extended in ways that are partly inspired by the architecture of the real-world Jane Street Exchange. DBSE relies fundamentally on UDP and TCP network communications protocols and implements a subset of the FIX *de facto* standard protocol for financial information exchange. We show results from an example in which the exchange server is remotely launched on a cloud facility located in London (UK), with trader clients running in Ohio (USA) and Sydney (Australia). We close with discussion of how our simulator could be further used to study planetary-scale latency arbitrage in financial markets.

## 1. INTRODUCTION AND MOTIVATION

### 1.1 Algorithmic Trading & Experimental Economics

Since the dawn of the first financial exchange – the Amsterdam Stock Exchange in 1602 – until the introduction of computer aided trading in the 1970s, the buying and selling of financial products, such as shares and bonds were executed by the verbose shouting of highly-paid individuals on the floors of major financial exchanges. By today's standards, these interactions between skilled human traders were slow, inefficient, and often error prone. Consequently, as soon as the relevant technologies were available, the buying and selling of financial products became a digital interaction and the traditional trading floors in national financial exchanges around the world were gradually closed. Modern-day electronic financial exchanges are highly sophisticated and complicated distributed computational systems that enable institutions, such as investment banks, fund managers, brokers, and insurance companies, to remotely connect and trade on the world's open markets. Although the transition from physical to electronic markets was largely complete (or, at least, inevitable) by the end of the 20[th] Century, at the turn of the millennium most trades were still executed by humans. Since then, as computer hardware capabilities improved and as regulatory barriers were lowered, the financial markets underwent a second technological revolution: the widespread introduction of automated software systems that replace human traders.

In many major present-day markets, the majority of trades are executed by sophisticated autonomous adaptive computational systems. These automated systems (known variously within the industry as *trading agents*, *algo traders*, or *robot traders*) can be responsible at any one major investment bank for weekly trading flows of $100Bn or more. Early automated trading systems failed to match the behavior of human traders, but in 2001 a research team at IBM's TJ Watson Labs published results (Das *et al.* 2001) from an experiment that tested the effectiveness of two adaptive trading algorithms, known as GD (Gjerstad & Dickhaut 1998) and ZIP (Cliff 1997). Using heterogeneous populations of humans and agents, the IBM researchers discovered that the GD and ZIP trading algorithms could consistently outperform human traders with respect both to efficiency and to profitability. In the past 15 years, the rising penetration of this technology has transformed the financial markets into a world of predominantly robotic traders, not necessarily a change for the better, as documented by Arnuk & Saluzzi (2012), Bodek & Dolgopolov (2015), Cartea *et al.* (2015), Narang (2013), Patterson (2013), and Rodgers (2016). For a

comprehensive introduction to exchanges and trading, see Harris (2002).

Robot traders are capable of processing vast amounts of data and can react to market changes at millisecond speeds. Consequently, trading agents are considered by some to be vastly superior to human traders because not only are they more profitable, but also their operating costs are often substantially cheaper. Now, rather than hiring lots of human traders, a trading house can instead clone robotic agents and deploy them to multiple servers, with the primary operational costs no longer being salaries and bonuses but instead being those associated with maintaining the trading software, and the server-hardware that it runs on. Many financial institutions and proprietary trading firms now focus solely on the research and development of more intelligent, faster, and more profitable advanced trading algorithms. However, investing in algorithms that can quickly make intelligent trading decisions is not all that is required: many trading houses also make major investments in reduction and mitigation of telecommunications latency. The reason for this concentration on latency is best illustrated by an example: if two competing, identical agents are listening to market events from the same exchange, the one which can receive and respond the quickest will be the most profitable, because its actions in the market may well require the slower agent to re-think, i.e. to re-start the process of deciding what to do next. Hence being the cleverest trader is not necessarily a guarantee of riches: if a slightly less clever, but much faster, trader is active in the market, that trader's sheer speed may be enough for it to outperform the greater brainpower of the cleverer trader. The industry in recent years has been focused on a battle to minimize latency between robotic agents and exchanges (see e.g. Haldane 2011). This enables trades to be fulfilled electronically at super-human speeds, potentially trading multiple times per second, a practice known as High Frequency Trading (HFT).

In conjunction with these real-world developments, academics have been exploring the profitability of both human and robotic traders with experiments and simulations for decades. In 2002, the Nobel Prize in Economics was awarded to Vernon Smith, for his ground-breaking work in establishing a new field of research now known as *Experimental Economics*. Starting in the late 1950's, Smith worked on a series of laboratory experiments which demonstrated that the behavior of human traders could be studied empirically under controlled and repeatable laboratory conditions: see e.g. Smith (1992), and Kagel & Roth (1997, 2016).

In 2012, a minimal simulation of a modern-day electronic financial market was released as free open-source software, as a tool for teaching and research on automated trading algorithms: this software is the *Bristol Stock Exchange* (BSE), documented in (BSE, 2012) and in (Cliff 2018a, 2018b). BSE allows experiments in the style pioneered by Smith to be run in simulation on a standard laptop or desktop computer. A simulated market requires virtual traders, and BSE includes source-code for a number of robot-trading strategies, including: *Giveaway* (Cliff, 2018a); *Zero-Intelligence-Constrained* (ZIC: see Gode & Sunder 1993); *Shaver* (Cliff, 2018a); *Sniper* (inspired by Kaplan's Sniper strategy described by Palmer *et al.* 1992); and *Zero-Intelligence-Plus* (ZIP: Cliff 1997). Within the BSE documentation, these strategies are referred to via three- or four-letter codes: GVWY; ZIC; SHVR; SNPR; and ZIP, respectively. Despite BSE's success as an introductory teaching aid, it has significant limitations compared to real-world distributed financial exchanges: the most notable of which, like many other financial exchange simulations, is that BSE is only a minimal approximation to a fully distributed real-time system, and therefore critically it assumes absolutely zero latency in the communications between trader and exchange. Yet, as noted above, minimizing the latency of robotic agents is an integral part of their design and operation, and thus simulating robot traders without also modelling latency is a major limitation to using BSE as a platform for leading-edge research. This paper addresses that limitation.

In the text that follows, we introduce a significantly extended and truly *distributed* version of BSE, referred to as DBSE. Like BSE, DBSE implements a financial exchange simulation that can be used with robotic agents; unlike BSE, DBSE allows for the creation of truly distributed experimental systems that operate asynchronously in real-time, and with real communications latencies. DBSE has been written as a cloud-native simulator: it has been designed from the outset so that constituents of DBSE can be launched in cloud data-centers around the world, enabling the user of DBSE to set up a truly planetary-scale distributed system consisting of an exchange server that provides connectivity to trading clients at various disparate locations on the planet. This allows for the study of mechanisms to deal with real-world communications latency, rather than simulated approximations of it. Practical use of DBSE is only possible because of the availability of low-cost remotely-accessible cloud-computing services: the example we give here (and for which we have released the code as open-source on GitHub) runs on Amazon Web Services but adapting it to any other major cloud provider would be trivial work. The ultimate output of the work reported here is our release of the DBSE code-base as open-source, that can be used by researchers and practitioners as a common platform for exploration of many other key questions within Experimental Economics.

### 1.2 Testing Robot Traders with Realistic Latency

A fundamental challenge with testing robot traders is answering the question of where you can test them. Running a prototype algorithm on a real-world financial exchange would be the most realistic test, however this carries a manifest risk of serious financial consequences if anything were to go wrong. Even if the losses were bearable, there are often significant regulatory barriers to also overcome. Consequently, it is preferable to test new agents in controlled simulations that accurately imitate real markets. Many organizations choose to use playback simulations, a process of replaying legacy real market data and recording how agents react to the changes in market events. These simulations are fundamentally flawed as the agent is not able to influence the change in market prices with its own activity: regardless of what the agent does, the prices on the simulated market immediately after the trader has sold or bought a large quantity of some asset will remain the same as if the trader had instead done nothing at all. The alternative is to build your own market simulator, one in which the actions of an individual trader really can have an immediate effect on the rest of the market, and the Bristol Stock Exchange (BSE) is an example of such a simulator.

BSE allows the user to configure how the supply and demand for a tradeable asset will change over time: this influences (but does not exclusively determine) the market price of that asset; critically, the actions of individual traders also influence prices on a moment-by-moment basis. The disadvantage however is that the trading agent is operating on data from a synthetic market. Whether a market simulator is based on legacy-data playback or on dynamically determined market prices, these simulations are very often designed to run on a single machine. In the case of BSE, the simulator runs in a single thread, with the operations of the exchange and all the active traders running via time-sliced simulation techniques operating in the application's sole thread. This results in absolute-zero latency between the traders and exchange because the agents continuously have perfect knowledge of the data broadcast by the exchange, and because their messages back to the exchange take exactly zero seconds to get there, significantly limiting the realism of the simulation. Other public-domain exchange/market simulators, such as De Luca's *Open Exchange* (De Luca & Cliff 2011; De Luca 2015) do run as distributed systems across a local-area network (LAN): in these there will be fundamentally non-zero latencies as network packets are communicated around the LAN, but at the level of the practitioners running the experiments these latencies are so small as to be treated as negligible, i.e. as effectively zero.

For these reasons, there is a strong requirement for better simulations, both in the academic and professional world: both academic researchers and professional industry practitioners care about latency not only as a source of frustrating delays, but also as an aspect of the market that could potentially be profitably exploited, via what is known as *latency arbitrage*.

Latency arbitrage involves exploiting time disparities in the market. The disparity may be between the public price of a stock and the latest market update, or it could be that one of several available trading venues (i.e., exchanges) is particularly faster or slower at processing and responding to orders than other venues, in which case there are ways of profitably exploiting the speed difference. High frequency trading (HFT) firms pay very large amounts of money for their computer systems to be as physically close to their target financial exchanges as possible and to have direct access to the market data publishing feeds. Latency arbitrage can be exploited when multiple exchanges are selling the same commodity, for example let's assume both the NASDAQ and the New York Stock Exchange (NYSE) are selling a stock XYZ. At time, $t=0$, the price of XYZ throughout America is constant. This is governed by the Securities Information Processor (SIP), which links all U.S. markets by processing and consolidating all quotes and trades from every trading venue into a single data feed (CTA, 2018). If a large trade was to occur on the NYSE, thus changing the price of XYZ on that exchange, market data would be published to both the SIP and along the direct data feeds to HFT firms. Due to the SIP consolidating market data, it is slightly slower than a direct feed to a HFT firm and so for a few fractions of a second, the price of XYZ on the NYSE is different to the NASDAQ. HFT firms can react to this disparity in price and buy or sell XYZ accordingly between the NYSE and the NASDAQ, before the NASDAQ even learns about the changed price from the SIP. Although perhaps a morally questionable practice, this is a technique that trading firms routinely use to make significant profits. Most market simulations, including BSE, have no concept of latency and thus it is impossible to simulate this style of trading. Understanding and developing techniques to minimize or exploit latency arbitrage is a constant demand in industry, and has also been the topic of academic research (see e.g. Wah & Wellman 2013; Duffin & Cartlidge 2018).

In this paper, which summarizes (Miles, 2019), we describe the design and construction of a distributed financial exchange that can test the profitability of trading agents in real-time with real latency. This work explores how real-world financial institutions and exchanges build their systems and communicate with each other across a distributed network for order placements, execution reports, and market data publications at scale, all whilst minimizing latency across the globe. Although our DBSE simulator was originally based on the publicly available BSE source-code, that code was extensively rewritten to provide more functionality to the executing agents, as well as removing the assumption of zero-latency. With suitable network configuration, users of this simulation will be able to remotely host a financial exchange at any location in the world where a cloud-hosting service is situated, and to connect numerous trading agents running on servers that are sited in regions that are geographically local or remote with respect to the server on which the exchange is running. Here we demonstrate the capabilities of this simulation deployed on the cloud compute and networking infrastructure provided by Amazon Web Services (AWS). Our work enables researchers and educators to design and evaluate new trading agents on a more realistic test-bed that utilizes the same communication technologies as real-world financial exchanges. DBSE has been designed and implemented with scalability in mind. This paper is our first description of DBSE, and there remain many opportunities to extend the project in various ways: we aim for the code-base to be accessible to all of those who wish to improve and work with it. The goal is to enhance the research and teaching capabilities of our software and for it to become a common platform: to the best of our knowledge DBSE is the first such globally distributable freely available open-source financial exchange simulation.

## 2. CASE STUDY: JANE STREET EXCHANGE

Running a successful electronic financial exchange can be a lucrative business, and technical details of commercially sensitive computer systems are rarely placed in the public domain. It is generally extremely difficult to find detailed information on the architectural designs of financial exchanges and even more challenging to find fine-grained descriptions of how trading firms organize their automated systems to place orders and use subscriptions to market data sources.

Nevertheless, on the 2nd February 2017, Jane Street Inc, a global liquidly provider and market-making company, published a technical presentation outlining a high-level overview of the design and development of their own financial exchange known as JX (Jane Street, 2017). Jane Street develop in-house proprietary models and use quantitative analysis to trade over $13 billion in equities worldwide in a single day. The motivation for the development of JX was due to their necessity to test new algorithms and financial models – much like the motivation for our work presented here. The JX exchange is based on the design of the American NASDAQ exchange and serves as a perfect exemplar for a real-world distributed financial exchange.

JX has been designed to satisfy many of the underlying requirements of a modern exchange, including dealing with high transaction rates, and aiming for consistently low response times, while maintaining fairness and reliability. It has been reported to handle messages rates in the 500k/second range with latencies in the single-digit microseconds. Below we summarize key roles and responsibilities of the individual micro-services on the JX internal private network and outline the key technologies that enable such high performance.

Figure 1 outlines the key components of the JX distributed limit order book financial exchange. The network backbone is represented by the thick purple line; any component above this line has no direct communication with the outside world and can be assumed to sit within a private subnet of the network.

All components below the network backbone are public facing and can connect to external clients either via the Internet or a paid private connection, at the exchange owner's discretion.

### 2.1. The Matching Engine (ME)

The matching engine (ME) is the heart of the exchange. It is a single, monolithic machine that holds all the current orders on the exchange in a Limit Order Book (LOB) data structure, as discussed by e.g. Harris (2002), Cartea *et al.* (2015), and Cliff (2018a, 2018b). These orders are kept in memory and the matching engine is responsible for identifying the buy and sell orders that can be matched to give each counterparty a transaction that satisfies their order. The JX Matching Engine sits within a strictly managed internal private network and receives and publishes message to a wide variety of other services on the client side of the network.

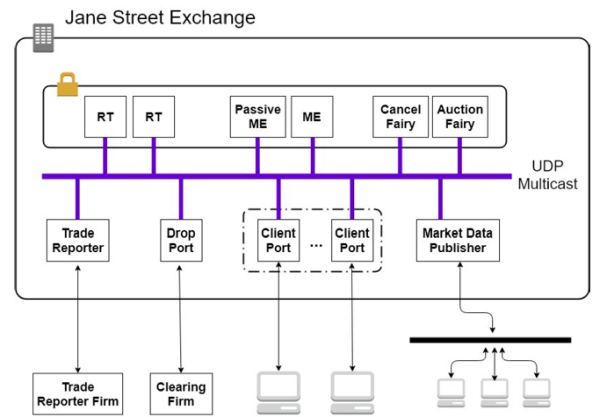

Figure 1: Architecture of the Jane Street Exchange.

Adjacent to the active Matching Engine is a passive copy that listens to all output of the active matching engine: this *Passive ME* is used as a fail-safe if the active engine has a system failure and goes offline, at which point the passive engine immediately takes control. In real-world exchanges this failover process is close to instantaneous to limit impact to clients and to the wider financial world.

### 2.2. Cancel Fairy & Auction Fairy

As previously stated, the ME is the critical component of a financial exchange and must losslessly deal with receiving thousands of messages a second. A large proportion of these messages however are not relevant to the live activities of the exchange; an example of which is the *cancel order* request message. At the end of a market session, it is common for clients to cancel all their live orders. Clients would send cancel order requests throughout the trading day but set a delay so that each of them is only executed at market close.

Historically, it was the responsibility of the ME to keep track of all these delayed cancel order requests. This however added noise to the matching engine, and thus modern exchanges offload this responsibility to an independent application, known as the *Cancel Fairy*. When a delayed cancel order request hits the ME, it would be acknowledged by the ME and then picked up by the Cancel Fairy. Once it is time for the order to be legitimately cancelled on the exchange, the Cancel Fairy would send a new cancel request that would be executed immediately by the ME.

In other non-continuous exchanges, there exists an additional process known as the *Auction Fairy*. It is used to aggregate many orders, often with overlapping prices, and runs an optimization to find a price that maximizes the shares traded. This process takes time and so is run independently. Once complete, the results are returned to the ME; another example of how modern distributed exchanges strive to limit the work done by the ME.

### 2.3. Re-Transmitter (RT)

In order to make communication across the network efficient and fair, all components communicate with each other via a technology known as UDP Multicast, discussed in more detail in Section 3.2. Critically this technology does not include guaranteed message delivery or acknowledgement and so messages can be lost during transmission. To account for this, a series of processes known as re-transmitters are added to the network. Their sole purpose it to record all the messages that have been seen on the network and to resend any message that was subsequently lost by a micro-service. Multiple re-transmitters are used in the event that one of them did not receive a message and thus they can communicate and reach consensus about the current state of the messages transmitted. In the unlikely event that all of the re-transmitters did not receive a message, they can request it from the ME, which maintains an in-memory copy of all messages sent in the current market day.

### 2.4. Client Port

Each Client Port is the main connection point for external financial institutions, such as brokers, investment banks, and fund managers wanting to trade on the exchange. The client ports accept connections to individual clients and provide a mechanism for them to perform transactions with the matching engine. This includes placing, amending and cancelling orders, requesting quotes and receiving execution reports on any trades that resulted from that client's orders.

### 2.5. Drop Port

The Drop Port is very similar to a Client Port, however it accepts connections from institutions known as clearing firms instead of clients. When a trade occurs, an independent third party is needed to take responsibility for "clearing" the trade, i.e. managing the transmission of money and transfer of ownership between the two trading parties. This is the responsibility of clearing firms, and thus they require information about the activities of both clients. As such, when a trade occurs and the execution reports are sent to the corresponding client ports; both execution reports are sent to and aggregated by the Drop Port and then sent to the relevant clearing firm.

### 2.6. Trade Reporter

The Trade Reporter is the public-facing data feed for all trade activities on the exchange. It listens to all trades that occurred on the matching engine, anonymizes the data, and publishes it to an external trade reporting firm.

### 2.7. Market Data Publisher

Similar to a trade reporter, a Market Data Publisher (MDP) listens to and anonymizes all market data on the matching engine. Instead of broadcasting to an external trade reporting facility however, the market data publisher uses UDP multicast technology to transmit the data to a network of clients, both human and robotic. The fees to access these MDPs are very expensive; this is how many HFT firms take advantage of latency arbitrage, using an MDP to know the exchange's activity before any competitors reliant on the Trade Reporter do.

## 3. DBSE IMPLEMENTATION DETAILS
### 3.1 FIX for Financial Information eXchange

A major design commitment was our decision to use the Financial Information eXchange (FIX) Protocol (see FIX, 1992) as DBSE's inter-process communication language for order placement and execution reports. FIX is unarguably the communication protocol of choice in real-world finance; it is used as a *de facto* standard by thousands of financial institutions and exchanges daily to facilitate trading data exchange. Understandably, for the FIX protocol to handle all aspects of financial trading in the real world, it supports a large and complex language of different messages. This is a noticeable disadvantage for its use within DBSE because FIX's messaging capabilities are far more extensive than what is required for DBSE in its current form. In the latest versions of FIX, the protocol supports messages for all aspects of stock trading as well as other financial asset classes, including bonds and foreign exchange.

As a result, it could be argued that the FIX protocol provides too much functionality that complicates the development of DBSE. Instead, a more simplistic protocol could have been used and the messaging language customized for DBSE's needs. The counterargument is that using FIX enriches DBSE as a teaching platform because it presents to users the real-world language and mechanisms that facilitate worldwide financial trading. Moreover, people who desire to do so can view the DBSE source-code to observe how the protocol operates and is implemented within the exchange, and in teaching contexts using DBSE provides an opportunity for students to experience creating FIX-compliant trading clients of their own as a potential coursework assignment. From a realism perspective, the decision to use FIX undoubtedly enhances DBSE. FIX is the global trading protocol and thus the time required to send FIX messages on the DBSE should be close if not equivalent to that of real-world financial institutions: although precise data on such timings is generally not publicly available. Use of FIX supports DBSE's overarching goal of being real-time and using real world tools wherever possible. Finally, regardless of whether FIX is too extensive or not, the selected protocol for DBSE had to fulfil three main characteristics. It had to be bi-directional, full-duplex, and able to communicate over a single constant TCP connection. Unsurprisingly, the FIX protocol supports all three of these characteristics, because it was designed to support financial communication, and that made it a natural choice.

## 3.2. UDP Unicast

Our final deployment of DBSE uses the UDP protocol with the unicast addressing method for publication of market data from the exchange to trading clients. This combination is close to, but not an exact copy of, what is used in the Jane Street Exchange and on other real-world financial exchanges. Ideally, the market data would be published using UDP multicast rather than unicast, to ensure efficient, non-duplicated traffic throughout the network. A compromise unfortunately had to be made because currently AWS does not support the multicast addressing method. Because of this, TCP was a consideration to replace UDP as it would guarantee message delivery. Upon detailed evaluation however, TCP would require the exchange to manage connections between all trading clients, increasing its computational overhead. Moreover, in the event of a lost packet during UDP transmission, it does not cause a major issue to clients as they will just update their market data when a future packet is received. Despite the compromise of using UDP unicast rather than multicast, DBSE still publishes market data successfully to clients positioned in cloud data centers across the globe. The only slight negative consequence is that the exchange's publisher must iterate though each client in turn sending them their market data. This does not cause any issues at the current scale of our DBSE deployments: in the next section we show DBSE supporting four trading clients at various locations around the world, with each trading client playing host to multiple robot traders. In any case, DBSE maintains an implementation of both unicast and multicast transmission so if at a later date AWS starts to support multicast, or a user wishes to buy and maintain their own networking hardware for larger-scale tests, the multicast functionality built into DBSE can be brought into use.

## 4. TESTS AND EVALUATIONS

### 4.1. Latency Tests

To demonstrate and evaluate the UDP unicast market data publisher and the distributed nature of DBSE, we conducted an investigation into the varying latencies for clients positioned around the globe. Specifically, with a DBSE exchange-server hosted in London, we timed how long it takes for trading clients in London, Ohio and Sydney to receive market data. We wanted to ensure that there was a disparity in latency depending on how far from the exchange a client was hosted. This was crucial because without distance-dependent variations in latency, it would be impossible to test whether the profitability of a trading agent is dependent on its ability to race-to-market. To test the latency, we ran DBSE with some additional timing code. When the exchange publishes market data it timestamps the message before sending it through the network. Thus, when each respective trading client receives the message it can perform a comparison between the time of arrival and the timestamp for when the message was sent: that timestamp is located within the message. By utilizing the network atomic clocks provided by the Amazon Time Sync Service we could guarantee that all time on the network would be synchronized and thus the results would be accurate. Results presented below are from an experiment that ran for ten minutes, recording the latency to transmit market data from a DBSE exchange-server in London to four trading clients under otherwise routine simulation conditions.

During the ten-minute experiment, DBSE published market data 491 times. Table 1 summarizes the minimum, first quartile, median, third quartile and maximum latency timings, in milliseconds, for each of the four clients. CLNT1 and CLNT2 are both located in the London, UK region; CLNT3 is positioned in Ohio, USA; and CLNT4 is hosted in Sydney, Australia. As expected, the results show that as geographical distance increases from the DBSE exchange server, so does the latency. Consequently, clients located in Australia receive market data from the exchange in a median time of 135.3ms compared to London's 0.8ms and 0.9ms. Interestingly, market data is consistently received by CLNT2 0.1ms slower than CLNT1, even though they are located within the same region. This is likely because of using the unicast addressing method instead of multicast, as a result of the exchange sending data to each client sequentially; thus, CLNT2 is usually sent data fractions of a millisecond after CLNT1.

|  | Latency (ms) | | | |
| --- | --- | --- | --- | --- |
|  | CLNT1 | CLNT2 | CLNT3 | CLNT4 |
| MIN | 0.4 | 0.5 | 43.7 | 134.9 |
| Q1 | 0.7 | 0.8 | 44.0 | 135.2 |
| MEDIAN | 0.8 | 0.9 | 44.1 | 135.3 |
| Q3 | 1.0 | 1.1 | 44.3 | 135.5 |
| MAX | 2.9 | 1.8 | 55.0 | 138.5 |

Table 1: Results for the latency experiment. The DBSE exchange server is located in London, as are clients CLNT1 and CLNT2. Clients CLNT3 and CLNT4 are in USE and Australia, respectively.

The spread of the latency timings is relatively consistent amongst all four clients, although there are a couple of outliers that result in the high maximum values of 2.9ms and 55ms for CLNT1 and CLNT3 respectively. Figure 2 shows the distributions of each client's latency, binned into 0.1ms intervals.

All four graphs in Figure 2 have approximately the same left-skewed shape and in each the majority of latency is clustered within a 0.5ms spread. Table 2 presents the mean, variance and standard deviation of the timing experiment. These results show that CLNT1, CLNT2 CLNT3 and CLNT4 each on average receive market data 0.9ms, 1.0ms, 44.2ms and 135.4ms after the exchange publishes it. This was to be expected, as transmitting messages over increasingly greater distances should take longer amounts of time.

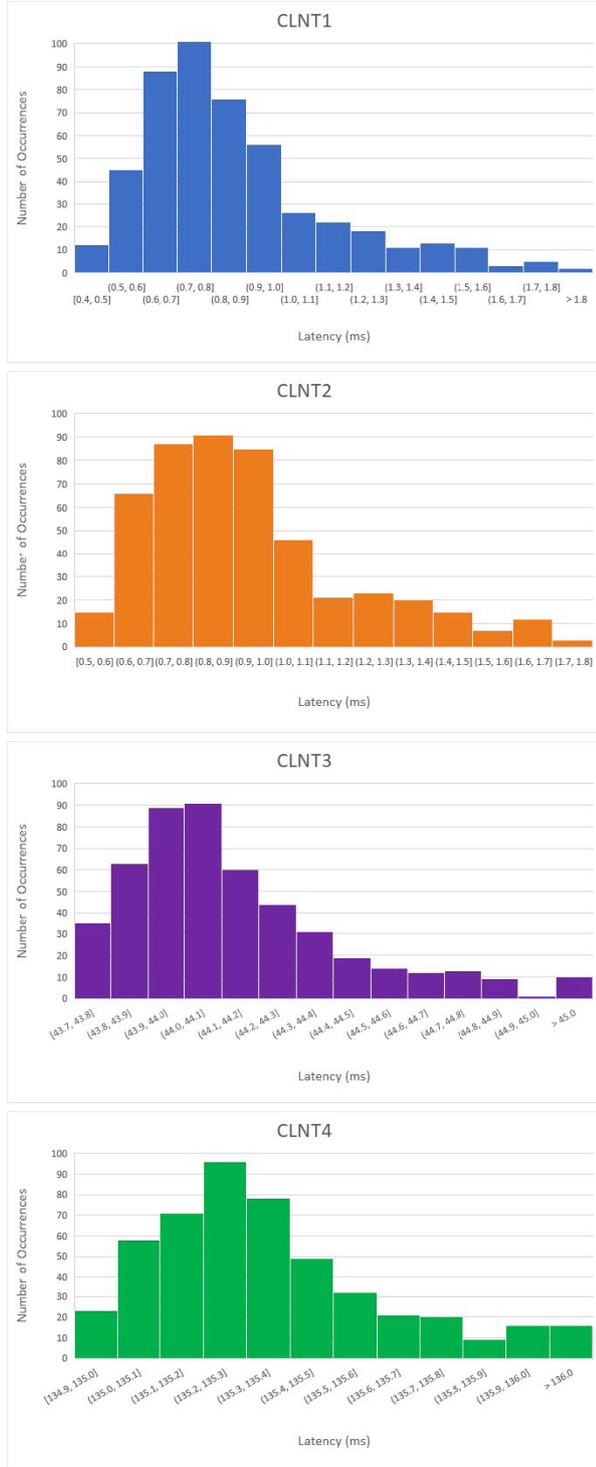

Figure 2: Latency distributions for four clients.

However, the values for the variance and standard deviation of CLNT3, positioned in America, were unexpected compared to the other clients. Since all communication traffic was occurring within AWS's internal network, we would have expected the variance and standard deviation of latency across clients to be consistent. Upon further analysis of the timing data, the larger spread of CLNT3 was caused because of a few outliers, the largest of which was 55ms. This gives insight into the amount of traffic AWS's internal network is handling between London and Ohio: these increased latencies suggest that Amazon handles more spikes in traffic between London and Ohio.

|  | Latency (ms) | | | |
| --- | --- | --- | --- | --- |
|  | CLNT1 | CLNT2 | CLNT3 | CLNT4 |
| MEAN | 0.9 | 1.0 | 44.2 | 135.4 |
| VARIANCE | 0.1 | 0.1 | 0.4 | 0.1 |
| STANDARD DEVIATION | 0.3 | 0.3 | 0.7 | 0.3 |

Table 2: Spread of the latency experiment.

These results demonstrate that UDP unicast definitely is a perfectly viable option for transmitting market data within Amazon's network to clients positioned across the globe. UDP was the logical choice, compared to TCP, as it is fast, requires little computational overhead, and is the protocol used by real world exchanges. Despite being restricted to the unicast addressing method, DBSE successfully handles it role at millisecond speeds with the current configuration of trading clients.

### 4.2. Race-to-Market Experiment

To demonstrate the capability of DBSE as a real-time and real-latency simulation we conducted a race-to-market experiment. As discussed earlier in this paper, race-to-market is a concept by which a trader can "steal the deal" if they learn about and respond to a market change before a competitor. Therefore, in a real-world scenario, if a trading client is positioned further away from the exchange than a competitor's trading client, then it will take longer for that client to receive market data. Consequently, the closer of the two clients can react faster to market events and therefore should be more profitable. We explored this in DBSE.

We constructed an experiment on the globally deployed DBSE with four configured trading clients, two in London, one in Ohio and one in Sydney. The experiment would consist of a total of 160 trading agents across the four clients. These trading agents were split 50/50 between supply and demand as well as 25/25/25/25 between four of BSE's built-in trading algorithms, Giveaway (GVWY), Shaver (SHVR), Sniper (SNPR) and Zero-Intelligence Constrained (ZIC). For each trading client, there were five agents of each robot type on the supply side and five agents of each robot type on the demand side, hence a total of 40 trading robots per client, and 160 agents for the simulation across four clients.

Each of the four trading clients were given equivalent order scheduling configuration that ran for a total of three minutes. The order schedulers were configured to distribute new orders to the traders, for them each to either buy or sell some number of units of the exchange's tradeable commodity, at 30 second intervals: inter-arrival times of orders were set to follow a Poisson distribution (this functionality is built-in to the original BSE, via BSE's drip-poisson update mode). Within each three-minute simulation, the range of prices for both the supply and the demand are configured to change every minute. Initially, at time $t$=0, the supply and demand are configured to sit in the range $1.00–$2.00; at time $t$=60, the range increases to $1.50–$2.50; before returning to the initial range, $1.00–$2.00, at time $t$=120. We set the parameter stepmode of each range to be fixed, this results in DBSE creating an even spread of orders across the price range, resulting in a theoretical equilibrium price $P_0$ of $1.50, $2.00 and $1.50 cents for each minute of the simulation respectively: if the market is functioning as would be expected, then transaction prices should converge to the relevant $P_0$ within each one-minute period. It is common practice in experimental economics to configure simulations in this way; changing the $P_0$ value at a set point in time via a shock change in the market's supply and demand, and transaction prices are expected to reflect the market adapting to each shock change; this is an accepted way to test the reactiveness of trading agents -- in the real world, transaction prices are constantly changing depending on the world's events. If the supply and demand curves of the simulation were configured to be constant then the $P_0$ value would also be static, and thus the market dynamics would be somewhat stale. The full simulation configuration for this experiment can be found in Appendix B of (Miles, 2019). We repeated the three-minute experiment ten times and for each run recorded the total profit of each trader type. Figure 3 shows the average profits per client for each type of robot trader over the ten runs. For this specific order scheduling configuration, the results show that the GVWY, SHVR and SNPR traders all performed roughly equivalently across clients, with the ZIC algorithm performing the poorest. These results show that regardless of distance from the exchange, each algorithm performs equivalently in each region compared to its counterparts.

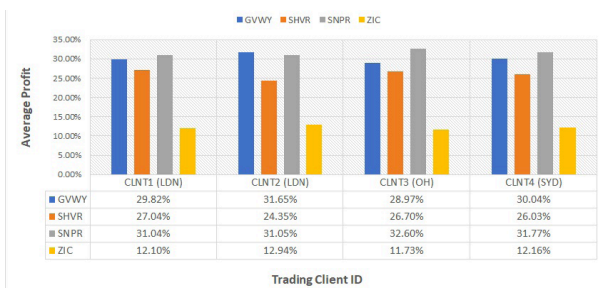

Figure 3: Ratios of total profit per trader type for each client.

Figure 4 on the other hand compares the total profits of all of the algorithms per client. The results presented here are particularly interesting as they indicate that on average CLNT1 and CLNT2 outperformed CLNT3, which in turn outperformed CLNT4. This supports our argument that increasing distance-related latency will degrade the performance trading agents of because CLNT1 and CLNT2 are positioned closest to exchange, followed by CLNT3, followed by CLNT4. Although the average profits of each client are close, there is a significant difference with CLNT2 in London earning 25.72% of profit compared to CLNT4 in Sydney earning 24.10% profit. If latency did not affect the profitability of trading agents and their ability to race-to-market, then we would have expected each client to perform equivalently and each earn 25% of profit across the simulation. These results show that latency can be a limiting factor in the profitability of agents. Designing new trading agents involves a challenging trade-off between adding more "intelligence" (which is typically more computationally demanding, in time and space) and keeping their total processing times low enough that the traders' reaction times keep them in contention in the race-to-market. The trading agents currently available in DBSE are all relatively computationally undemanding. Further work, discussed in Section 5, can be devoted to testing more sophisticated trading agents such as AA (Vytelingum, 2006), GDX (Tesauro & Bredin, 2002) or ZIP60 (Cliff, 2009) to determine whether the computational demands of their extra intelligence comes at the cost of their reactiveness to market events.

The results presented in this section have demonstrated that there is much to explore about algorithmic trading when one has access to a simulator that can offer real-time and real-latency analysis. DBSE enables such analysis and can be configured to enable researchers to uncover new insights into latency driven simulations.

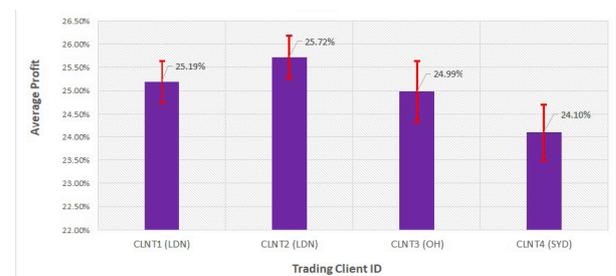

Figure 4: Ratios of total profit across clients.

## 5. FURTHER WORK

The ultimate aim of the work described here is to develop a distributed simulation platform that could fully model current multi-venue trading systems, and the opportunities for latency arbitrage between different venues. This would require extensive work, implementing multiple exchanges, a trade reporting facility between exchanges, and an entirely new trading client that could connect to and place orders on multiple exchanges simultaneously. Included in this work would be an expansion of the FIX messages that the current

DBSE exchange supports, such as the *Order Replace Request*, `<G>`, used to amend orders that are live on the exchange. Moreover, a persistent storage mechanism, such as a relational database, would benefit the exchange enabling it to be hosted permanently in the cloud. Such additional work could potentially consume many person-months of concentrated effort.

As a part of any future work, we propose a new high-level AWS architecture, as shown in Figure 5. This diagram does not include networking infrastructure but shows the simulator's compute hardware and introduces a new proposed application, the web client. Currently, it is inconvenient for users of DBSE to be required to SSH onto the simulation's hardware to run experiments. The web client would be a web-based application that acts as a simulation controller, hosted permanently in the cloud, that has the permission to orchestrate the instantiation, termination and synchronization of trading clients across the network. Protected behind a user access control system, such as that provided by Amazon Cognito (AWS, 2019), the web client would enable easy and efficient configuration of simulation runs in a graphical interface. Upon completion of a simulation session, it would amalgamate the results, terminate the unneeded trading clients, and provide suitable tools for analyzing the results.

Another obvious avenue for future work, already touched upon earlier in this paper, is the addition of more sophisticated automated trading agents implementing various of the strategies that have been described in public-domain literature, such as: AA (Vytelingum 2006, Vytelingum *et al.* 2008); ASAD (Stotter *et al.* 2013); GDX (Tesauro & Bredin 2002); HBL (Gjerstad, 2003); MGD (Tsauro & Das 2001); Roth-Erev (e.g. Pentapalli, 2008); and ZIP60 (Cliff, 2009).

DBSE has the potential to be an easy-to-use simulation for non-developers, both in the academic and business worlds, and we are intrigued to see how it is developed and used by the wider community in the future.

## 6. CONCLUSION

The Distributed Bristol Stock Exchange (DBSE) is a globally distributable financial exchange simulation for research and teaching. Its source-code consists of two independent applications, `dbse_exchange` and `dbse_trading-client`, both available for download from:

- `github.com/bradleymiles17/dbse_exchange`
- `github.com/bradleymiles17/dbse_trading-client`

The codebase has been written in Python 3.7 (currently the latest version of this programming language) and all function/method declarations have been typed to assist readability for new users of the DBSE. Both applications use an argument parser when executing, and when attempting to run the application a user can view the required and optional parameters via the help, `-h`, flag.

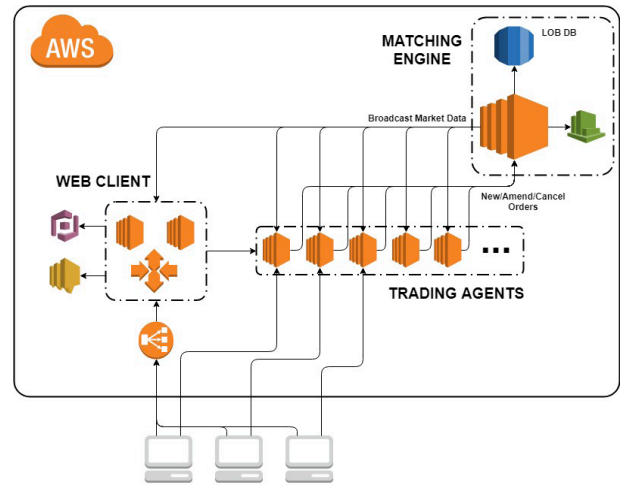

Figure 5: Proposed AWS architecture for future DBSE.

DBSE is significantly revised and expanded from the original work described in (BSE, 2012; Cliff 2018a, 2018b): it has extended the concepts embodied in the original Bristol Stock Exchange and taken BSE from a single-source-file single-threaded application into a fully distributed and cloud-native simulation that can readily be run on widely available commercial cloud-computing services. Trading clients can be configured and positioned around the globe and set trading simultaneously on a single stock exchange. Where BSE naively assumed absolutely zero-latency, DBSE operates using real-world financial communication protocols that are designed to minimize latency but which do not disregard it, and can be distributed at planetary scale for unavoidable real-world latencies. The results presented here demonstrate DBSE's capability in enabling research aimed at understanding race-to-market trading. DBSE is now offered to the global community of researchers and practitioners as a common platform for further exploration and tuition in how latency affects trading in contemporary markets, and in particular DBSE enables repeatable planetary-scale studies of latency arbitrage, a heavily under-researched topic in financial trading; it also serves as an open-source exemplar for teaching distributed systems architecture and design. We look forward to watching how the community makes use of this platform.